\documentclass{llncs}

\usepackage{makeidx}
\usepackage{amssymb}
\usepackage{booktabs}
\usepackage{hyperref}
\usepackage{url}
\usepackage{xcolor}
\usepackage{soul}

\begin{document}

\pagestyle{headings}

\title{Biomedical Question Answering via Weighted Neural Network Passage Retrieval}

\author{Ferenc Galk\'{o} and Carsten Eickhoff}

\authorrunning{Galk\'{o} and Eickhoff}

\institute{Department of Computer Science,\\ETH Zurich,\\Switzerland\\firstname.lastname@inf.ethz.ch}




\maketitle 

\begin{abstract}
The amount of publicly available biomedical literature has been growing rapidly in recent years, yet question answering systems still struggle to exploit the full potential of this source of data. In a preliminary processing step, many question answering systems rely on retrieval models for identifying relevant documents and passages. This paper proposes a weighted cosine distance retrieval scheme based on neural network word embeddings. Our experiments are based on publicly available data and tasks from the BioASQ biomedical question answering challenge and demonstrate significant performance gains  over a wide range of state-of-the-art models.
\end{abstract}
\section{Introduction}

Biomedical question answering is a key task in catering to clinical decision support and personal health information needs. Finding useful information in the extensive collections of scholarly biomedical articles poses a challenge to highly trained practitioners and patients alike~\cite{malakasiotis2014challenge}. Biomedical questions phrased by experts are usually more specific than common Web search queries, making them difficult to satisfy using general-purpose retrieval models. A typical expert question (for
example ``\textit{Which enzymes synthesize catecholamines in adrenal glands?}'') shows specific domain knowledge and aims at more than just a general overview about adrenal glands~\cite{tsatsaronis2015overview}. Instead, the expert is looking for a narrow set of documents or snippets that help answering a specific relation of the queried entities. Since many question answering systems rely heavily on document and passage retrieval, an increased performance in these tasks tends to propagate into significant QA performance gains~\cite{schulze2016hpi,voorhees2001trec}. While the task of whole-document retrieval for QA is well understood, passage-level retrieval approaches have received comparably less attention~\cite{lee2016ksanswer,nyberg2016learning,papagiannopoulou2016large,schulze2016hpi}.

This paper describes a piece of work in progress, focusing on passage-level retrieval for biomedical question answering on the basis of weighted combinations of neural network word embeddings. It makes the following contributions: (1) We propose a novel approach for weighting query embedding vectors for the cosine distance text matching scheme and show that it outperforms traditional models. (2) We demonstrate significant performance improvements over state-of-the-art neural network retrieval models on a sizable publicly available benchmarking dataset.

The remainder of this paper is structured as follows: 
Section~\ref{sec:method} formally introduces the problem domain as well as the proposed method. Section~\ref{sec:experiments} empirically evaluates the merit of our method on the basis of publicly available data and tasks created in the context of the BioAsq biomedical question answering challenge. Finally, Section~\ref{sec:conclusion} concludes with a summary and an outlook on future directions.


\section{Method}\label{sec:method}

Let us assume a textual question $Q$ consisting of individual terms, where each term is represented by a fixed-length vector. To satisfy $Q$, we rely on a large collection of biomedical documents organized in passages. In order to represent and retain the semantic content of the terms, we project them into semantically meaningful vector spaces, such as induced by Word2Vec~\cite{mikolov2013distributed} or GloVe~\cite{pennington2014glove}. In this way both the question $Q = q_1, q_2, \dots, q_i$ and each passage $P = p_1, p_2, \dots, p_j$ can be represented as a sequence of fixed-length vectors. In this high-dimensional space, we can now relate individual terms to each other by means of distance functions. A commonly employed measure of relatedness is the cosine of the angle between word vectors as expressed by the cosine distance.

$$cos(q,p) = 1- \frac{qp}{\|q\|_2 \|p\|_2}$$

Given such a representation as well as measure of similarity, we can now perform a variety of operations ranging from clustering to retrieval~\cite{brokos2016using,huang2008similarity}. In order to represent multi-term units such as entire queries and passages, several authors recommend uniform averaging of word vectors before finally returning a ranked top-$k$ list of closest passages to the query.

$$CD(Q,P) = 1- \frac{(\frac{1}{|Q|}\sum q_i)(\frac{1}{|P|} \sum p_j)}{\|\frac{1}{|Q|}\sum q_i\|_2 \|\frac{1}{|P|}\sum p_j\|_2}$$

\subsection{Weighted cosine distance}

One serious issue with the well-known cosine distance retrieval approach introduced above is that the uniform average vector is a poor semantic representation for general texts. This is due to the presence and abundance of stop words such as ``\textit{the}'', ``\textit{a}'', or ``\textit{is}'' that carry little meaning. By assigning uniform weights to all words, we water down the semantic content of informative words such as ``\textit{neurodegenerative}''. To address this problem, non-uniform weighting schemes such as idf can be used. A term's collection-wide inverse document frequency (idf) captures its uniqueness and assigns larger weights to rarer words on a logarithmic scale. By assigning idf instead of uniform weights to the words, a substantial increase in performance can be achieved as we will show in Section~\ref{sec:experiments}. 

$$CD_{idf}(Q,P) = 1- \frac{(\frac{1}{\sum idf(q_i)}\sum idf(q_i) q_i)(\frac{1}{\sum idf(p_j)} \sum idf(p_j) p_j)}{\|\frac{1}{\sum idf(q_i)}\sum idf(q_i) q_i\|_2 \|\frac{1}{\sum idf(p_j)}\sum idf(p_j) p_j\|_2}$$

While there are alternative approaches for aggregation such as position encoding~\cite{sukhbaatar2015end}, that can take into consideration the ordering of words, a series of preliminary studies suggest that idf weights perform best on the biomedical QA task. For the sake of brevity, we do not include these experiments in our empirical performance evaluation.

\subsection{Adjusted idf weights}
The previously presented idf weights depend solely on the distribution of words in the corpus. Incoming queries, on the other hand, may originate from a different distribution. Using idf scores generated from the document collection to weight query terms might result in a poor representation, since words that appear rarely in documents but frequently in questions can receive an unduly large weight. Question words such as ``\textit{what}'', ``\textit{when}'', ``\textit{where}'' are intuitive examples. When idf weights are calculated on a sizable sample of scholarly biomedical articles obtained from PubMed, the weight for ``\textit{what}'' is 5.05, whereas ``\textit{disease}'' (2.68), ``\textit{protein}'' (2.59) and ``\textit{artery}'' (4.16) end up being much less important despite their greater \textit{de facto} informativeness. 

As a consequence, passages containing the word ``\textit{what}'' will be estimated to be more similar to the question ``\textit{What is a degenerate protein?}'', according to our metric, than passages containing the more promising phrase ``\textit{degenerate protein}''. In fact, if we were to perform idf-based stopping and only retain the most important components of a question, our example question ``\textit{What is a degenerate protein?}'' becomes ``What degenerate?'' with the low-idf component ``\textit{is}'', ``\textit{a}''  and ``\textit{protein}'' being removed while the desired reduction in this case may have been ``\textit{degenerate protein?}'' which captures the topical essence of the original question much better.

We address this issue by generating idf weights from an alternative collection, specifically a mixture of large-scale corpora of biomedical~\cite{tsatsaronis2015overview} and general questions~\cite{rajpurkar2016squad}. Generating idf weights from the combined question set results in smaller
weights for general terms such as ``\textit{what}'', ``\textit{which}'', ``\textit{when}'' and 
larger weights for rarer, domain-specific terminology such as ``\textit{protein}'', ``\textit{disease}'', or ``\textit{artery}'', making it possible to capture the true intention of bio-medical questions and passages. We refer to this method as $CD_{q}$.

\section{Experiments}\label{sec:experiments}

Our empirical performance evaluation is based on documents and questions from the BioASQ 2017 challenge's document and snippet retrieval tasks \cite{bioasq2017}.  The goal in this task is to return the 10 most relevant passages from a collection of 12.8M PubMed abstracts for a specific biomedical question. A training set of 1799 manually curated questions along with relevant passages are provided by the challenge organizers. The test set is comprised of a separate set of 500 questions organized in five equally-sized batches. 

\subsection{Baselines}

To allow for a meaningful system comparison, we include a broad range of traditional as well as state-of-the-art performance baselines, trained and evaluated on the same data and task as our proposed cosine-distance based methods.

\begin{description}
\item[RND] This approach returns a random passage from the reference set of highly-ranked documents to create a weak baseline.
\item[MLP] This approach uses position-encoded sentence embeddings that are concatenated and eventually classified in binary fashion in a multi-layer perceptron~\cite{lee2016ksanswer}. This is a re-implementation of a system participating in a previous BioASQ challenge.
\item[MP] The Match Pyramid~\cite{matchpyramid2016} model generates a similarity matrix from the pairwise word interactions between question and candidate passage. We use a fixed 30 by 30 matrix with zero padding and the identity function as a measure of similarity. In a second step, convolutional filters condense the interaction matrix into a final vector representation for classification.
\item[DRMM] The Deep Relevance Matching Model~\cite{guo2016deep} is based on a similar scheme. It computes pair-wise term interactions between question and candidate passage that are then flattened into fixed-size histograms and  discretized and weighted to give the final vector representation.
\end{description}

\subsection{Experimental Setup}
All (machine learning) methods are trained using five-fold cross validation on the training set. Word embeddings for all methods are computed as length-50 word2vec vectors on the PubMed document corpus \cite{bioasq2017}. For each question (training and test), a reference set of highly ranked documents is given by the challenge organizers. We split these documents into individual sentences that will serve as our retrieval unit. Negative training examples for machine learning methods are randomly sampled from arbitrary non-relevant documents.

\subsection{Results}

Table~\ref{table:results} compares the various baselines and cosine distance variants in terms of mean average precision (MAP), Precision, Recall and $F_1$-scores each at a cut-off rank of 10 retrieved passages. Statistical significance of method differences is determined using a Wilcoxon signed-rank test at $\alpha < 0.05$-level and significant improvements over all baselines are indicated with an asterisk. While questions were originally grouped in batches, here we forego this structure in the interest of brevity. Batch-level scores showed some variance but displayed the same relative method ranking as the aggregate overview.

  \begin{table}
    \centering
    \caption{Passage retrieval performance for biomedical question answering.\label{table:results}}
    \centerline{
    \begin{tabular}{ l l l l l }
    \hline
      Method & MAP & Precision & Recall & $F_1$ \\
      \hline
      RND & 0.190 & 0.190 & 0.289 & 0.229\\
      MLP~\cite{lee2016ksanswer} & 0.226 & 0.236 & 0.352 & 0.282\\
      MP~\cite{matchpyramid2016} & 0.344 & 0.323 & 0.470 & 0.383\\
      DRMM~\cite{guo2016deep} & 0.348 & 0.344 & 0.510 & 0.411\\
      $CD$ & 0.341 & 0.339 & 0.484 & 0.399\\
      $CD_{idf}$ & 0.344 & 0.348 & 0.487 & 0.406\\
      $CD_{q}$ & 0.377$^*$ & 0.374$^*$ & 0.519 & 0.434$^*$\\
      \hline
    \end{tabular}}
  \end{table}

Due to the limited length of the reference document list as well as each individual abstract, random sentence selection does surprisingly well and sets a lower limit to method performance. All compared approaches yield meaningful results and significantly outperform this baseline. MLP is the weakest machine learning approach with substantially lower performance scores than those achieved by MP and DRMM. Cosine distance rankings are clearly improved by idf term weighting, lifting their results on a level comparable to that of MP and DRMM. Our adjusted term-weighting scheme following statistics of a separate, more representative, question corpus introduces another improvement in result quality, leading to the best overall results and a significant improvement over all contesting methods. 

\section{Conclusion}\label{sec:conclusion}
This paper presents a piece of work in progress towards passage retrieval for biomedical question answering via weighted cosine distances. In place of highly parametric end-to-end ranking networks, we devise a number of lean non-parametric weighting schemes that account for the differences in term distribution between document and question corpora. Our experiments on publicly available BioASQ data demonstrate significant improvements over a range of ranking networks. Especially in academic settings where datasets are often not sufficiently large to robustly fit multitudes of neural network parameters, such light-weight architectures are of increased interest.

While we noted the significant difference in term distributions between corpora at a biomedical example, the solution, as such, is not specific to the biomedical domain. In the future, we aim to investigate more formally rigorous ways of accounting for such differences as well as to evaluate them on a wider range of topical domains.

\bibliographystyle{plain}
\bibliography{ref.bib}

\end{document}